\documentclass[twocolumn,prb,floatfix,citeautoscript,nofootinbib,superscriptaddress]{revtex4}
\usepackage{amsbsy}
\usepackage{latexsym,epsfig,graphicx}
\usepackage{dcolumn}
\usepackage{graphicx}
\usepackage{subfigure}
\usepackage{comment}
\usepackage{color}
\usepackage{bm}
\usepackage{mathrsfs}
\usepackage{amsfonts}
\usepackage{amsmath}
\usepackage{color}
\usepackage{amssymb}
\usepackage{xspace}
\usepackage{epstopdf}
\usepackage{tabularx}
\usepackage{longtable}
\usepackage[colorlinks=true, letterpaper=true, pdfstartview=FitV, linkcolor=blue, citecolor=blue, urlcolor=blue]{hyperref}
\usepackage[normalem]{ulem}
\usepackage{multirow}
\usepackage{esint}
\usepackage{mathdots}

\pdfoutput=1
\begin{document}

\title{Quantum spin-tensor Hall effect protected by pseudo time-reversal symmetry}
\author{Ya-Jie Wu}
\email{wuyajie@xatu.edu.cn}
\affiliation{School of Sciences, Xi'an Technological University, Xi'an 710032, China}
\author{Tong Li}
\affiliation{School of Sciences, Xi'an Technological University, Xi'an 710032, China}
\author{Junpeng Hou}
\email{jhou@pinterest.com}
\affiliation{Pinterest Inc., San Francisco, California 94103, USA}

\begin{abstract}
  The celebrated family of the Hall effect plays a fundamental role in modern physics. Starting from the anomalous Hall effect (AHE) and the quantum AHE (QAHE) with broken time-reversal symmetry (TRS) to their spinful generalizations, including spin Hall effect (SHE) and quantum SHE (QSHE) protected by TRS, they reveal rich transport and topological phenomena.  However, in larger-spin $S$ ($S>1/2$) systems, besides charge current and spin current, there arise higher-rank spin-tensor currents. Recent work has uncovered an interesting spin-tensor Hall effect with spin-tensor currents in these larger-spin systems. Taking a step further, this work discovers a new class of topological states of matter dubbed \textit{quantum spin-tensor Hall} (QSTH) insulators with broken TRS, and their nontrivial topology is protected by a unique \textit{pseudo-TRS}. Most strikingly, QSTH insulators exhibit a quantized rank-2 spin-tensor Hall conductivity, whereas both charge (rank-0) and spin (rank-1) conductivities vanish. We also fully characterize their topological properties and highlight the physical interpretations via the underlying connections to QSHE. Our work enriches the family of the famous Hall effects and sheds light on the intriguing topological state of matter in larger-spin systems. It further offers new avenues toward spin-tensor-tronics and low-power atomtronics.
\end{abstract}

\maketitle

\section{Introduction and motivation}

\begin{figure}	\includegraphics[width=0.98\columnwidth]{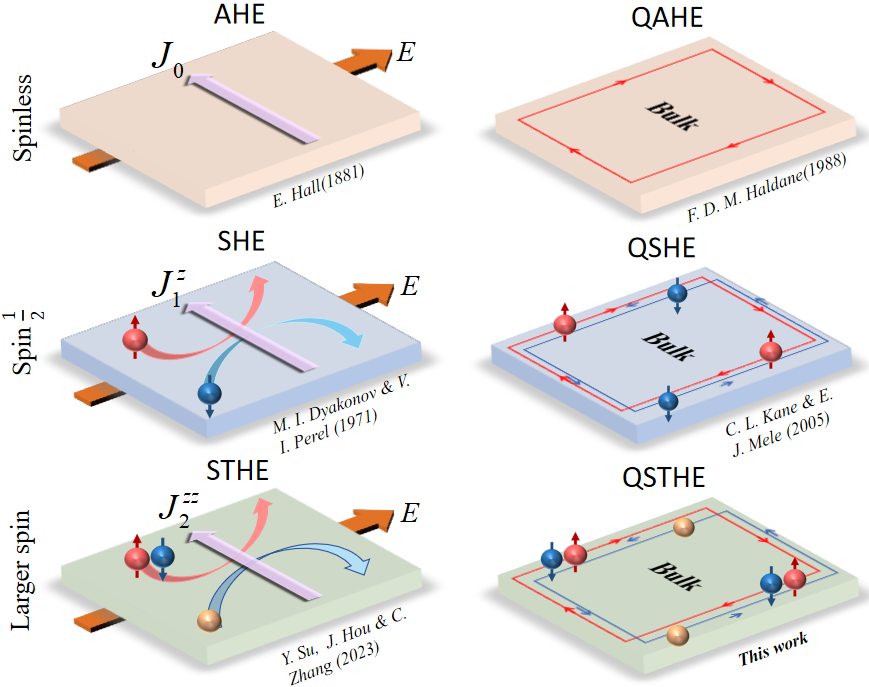} \caption{
The family of Hall effects.
     From left to right, we have the classical and quantum versions. From top to bottom, we show how they generalize along the spin degrees of freedom. This work completes the puzzle by introducing quantum spin-tensor Hall effect (QSTHE) on the bottom right.}
    \label{fig:hall}
 \end{figure}

The family of Hall effects (see Fig.~\ref{fig:hall}) has played a fundamental role in revealing rich topological and transport phenomena, representing a hallmark of modern physics. To start with, the anomalous Hall effect (AHE) arises from the interplay of spin-orbit coupling and broken time-reversal symmetry (TRS), leading to a transverse charge current $J_0$ even in the absence of an external magnetic field \cite{1,2}. Its quantized counterpart, the quantum anomalous Hall effect (QAHE), manifests in topological insulators with magnetization, hosting dissipationless chiral edge states \cite{3,444}. Enriched from AHE with the electron spins, the spin Hall effect (SHE) generates a transverse spin current $J_1^z$ without net charge flows $J_0$, driven by spin-orbit interactions \cite{5,6,7,8}, while its quantum version, the quantum spin Hall effect (QSHE), realizes a topological insulating phase with helical edge states protected by TRS \cite{9,10,11,12a,12} and has been experimentally observed in HgTe quantum wells \cite{13,14}. The Hall effect family not only deepens our understanding of topological states of matter but also holds great promise for low-power spintronic devices \cite{15,16,17,18,19,20} and quantum information applications \cite{PrivmanQuantum1998, LuFractional2024}.

Recent advances in simulating quantum phenomena using cold atoms have offered a tunable platform for studying many unique quantum states \cite{c2,c3,c4,c5,c6}. Most importantly, it enables the study of larger-spin systems, which leads to many intriguing topics, including spin-tensor-momentum coupling and exotic topological states \cite{l1,l2,l3,l4,l5,l7,l8,l9,l10,l11,l12}. Moreover, in larger-spin $S>1/2$ systems, the higher-rank spin-tensor current arises besides conventional charge and spin currents in electronic systems. Recently, in continuous space, a universal intrinsic higher-rank spin-tensor Hall effect (STHE) has been proposed in pseudospin-1 ultracold fermionic atoms beyond the scope of the conventional SHE \cite{sthe}. Interestingly, STHE induces a 
transverse higher-rank spin-tensor current $J_2^{zz}$ driven by a longitudinal external electric field.
A natural question arises: do physical laws promise a quantized STHE or a \textit{quantum spin-tensor Hall} (QSTH) insulator with only a quantized higher-rank spin-tensor current? If so, what's the symmetry protection, and how do we characterize its edge states and topological properties?

This work addresses these crucial questions by presenting the nontrivial lattice models that realize QSTH insulators. Our main results are summarized as follows:

(i) We construct a pseudospin-1 model on a honeycomb lattice for QSTH insulators with broken TRS, but it is protected by the \textit{pseudo-TRS} (pTRS) defined by a rank-2 spin tensor.

(ii) We examine the zigzag boundary states and the corresponding $\mathbb{Z}_2$ invariants with a focus on the spin compositions of the symmetry-protected edge states exhibiting rank-2 STHEs.

(iii) To validate the bulk-edge correspondence in QSTH insulators, we compute the rank-2 spin-tensor Hall conductivity using the Kubo formula directly. The results confirm a universal constant conductivity independent of the detailed model parameters. Meanwhile, both rank-0 charge and rank-1 spin Hall conductivities are zero, manifesting the unique QSTH phase.

(iv) We construct another toy model on a square lattice to indicate the versatility of topological QSTH insulators. Last but not least, we provide an intuitive physical interpretation by revealing the underlying connections between QSTH insulators and quantum spin Hall (QSH) insulators.

\section{Model Hamiltonian}
\begin{figure}	\includegraphics[width=0.9\columnwidth]{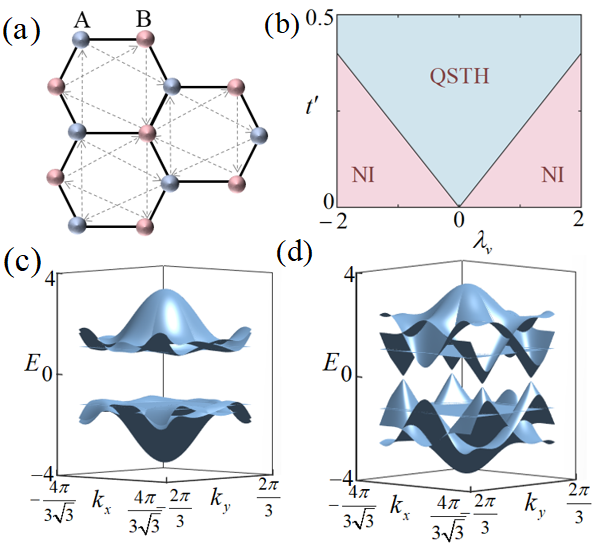}
 	\caption{
           (a) Illustration of the honeycomb lattice with A/B sublattices. Solid lines and dashed arrows correspond to the nearest-neighbor $t$ and next-nearest-neighbor $t'$ hopping. When the particles hop around the dashed arrows, the particles with pseudospin $\tau  = \pm 1$ acquire an accumulated phase $\pi/2$, and the particles with pseudospin $\tau  =0$ acquire an accumulated phase $-\pi/2$.
            (b) The phase diagram concerning $t'$ and the staggered sublattice potential $\lambda_v$. NI is a trivial normal insulator, and QSTH indicates the quantum spin tensor Hall state. 
            (c) and (d) The representative energy spectra in QSTH (${\lambda _v} = 0.2$) and NI (${\lambda _v} = 1.2$) phases, when $t'=0.2$. We set $u=1$ and choose $t=1$ as the unit of energy.
        \label{fig_band}
     }
 \end{figure}

We start with the following tight-binding Hamiltonian on a honeycomb lattice as in Fig.~\ref{fig_band}(a),
\begin{eqnarray} 
  H &=& t\sum\limits_{\left\langle {i,j} \right\rangle } {\hat{c}_{i}^\dagger } {s_1 }{\hat{c}_{j}}+ t'\sum\limits_{\left\langle {\left\langle {i,j} \right\rangle } \right\rangle} {{e^{i{\phi _{ij}}}}} \hat{c}_{i }^\dagger {s_2 }{\hat{c}_{j}}  \notag \\
 &+& \sum\limits_{i } {{\zeta _i}\hat{c}_{i }^\dagger \left({\lambda _v}s_1+us_3\right){\hat{c}_{i}}}
\label{Eq1}
\end{eqnarray}
under the basis $\hat{c}_{i}^\dagger =(\hat{c}_{i,-1}^\dagger ~ \hat{c}_{i,0}^\dagger ~ \hat{c}_{i,1}^\dagger)$, and $\hat{c}_{i,\tau}^\dagger$ is the creation operator of fermions with pseudospin $\tau  = \pm 1, 0$ on the $i$-th lattice site. Here $s_1 = {\bf{F}}^2 - 2F_z^2 + s_2 $, $s_2 = -N_{yy} + N_{zz}$, $s_3 = \frac{1}{2}(-{\bf{F}}^2+4F_z^2) - s_2 $, where ${\bf{F}}=(F_x, F_y, F_z)$ denote the rank-1 spin vectors, and ${N_{ij}}$ are rank-2 spin tensors defined by their anticommutator ${N_{ij}} = {\left\{ {{F_i},{F_j}} \right\}_ + }/2 - {\delta _{ij}}{{\bf{F}}^2}/3$ (see Appendix \ref{AppA} for more details). 
The strengths of the nearest-neighbor (NN)  and next-nearest-neighbor (NNN) hopping are denoted by $t$ and $t'$, respectively. The fermions accumulate a positive phase term $\pi/2$ when they hop clockwise on the honeycomb plaquette, as exhibited in Fig.~\ref{fig_band}(a). The last term in Eq.~(\ref{Eq1}) describes a staggered sublattice potential with $\zeta _i =1$ on A sites and $-1$ on B sites. This paper assumes a fermionic system that can be realized, e.g., via the hyperfine states of cold atoms \cite{f1, f2}. For simplicity, we refer to pseudospin as spin thereafter, unless otherwise specified.
 
Under the Fourier transformations $\hat c_{i \in A,\tau }^\dagger  = \frac{1}{{\sqrt {{N_s}} }}\sum\nolimits_k {{e^{ik \cdot i}}\hat a_{k,\tau }^\dagger } $ and $\hat c_{i \in B,\tau }^\dagger  = \frac{1}{{\sqrt {{N_s}} }}\sum\nolimits_k {{e^{ik \cdot i}}\hat b_{k,\tau }^\dagger } $ with $N_s$ the number of unit cells, the Hamiltonian ${\hat H}$ in Eq. (\ref{Eq1}) is given by ${{\hat H}} = \sum\nolimits_k {\hat C_k^\dagger h\left( k \right)} {{\hat C}_k}$ in the momentum space under the basis $\hat C_k^\dagger  = \left( {\hat a_{k,1}^\dagger ,\hat b_{k,1}^\dagger ,\hat a_{k,0}^\dagger ,\hat b_{k,0}^\dagger ,\hat a_{k, - 1}^\dagger ,\hat b_{k, - 1}^\dagger } \right)$. Here, the Hamiltonian matrix $h(k)$ reads
\begin{eqnarray} \label{Eq2}
h(k) &=& \left( {{\alpha _{R,k}}} {\sigma _x} + {{\alpha _{I,k}}}  {\sigma _y} + {\lambda _v} {\sigma _z}\right){s_1}  \notag \\
&+& \beta _k{\sigma _z}s_2+u {\sigma _z}s_3, 
\end{eqnarray}
where ${\alpha _{R,k}} $ and ${\alpha _{I,k}} $ refer to the real and imaginary parts of ${\alpha _k} = t\sum\nolimits_{i = 1}^3 {{e^{i{\rm{k}} \cdot {\delta _i}}}} $, respectively, with the NN vectors ${\delta _1} = \left( {1/2,\sqrt 3 /2} \right),{\delta _2} = \left( {1/2, - \sqrt 3 /2} \right)$ and ${\delta _3} = \left( { - 1,0} \right)$. The NNN hopping gives ${\beta _k} = 2t'\left( {\sin \sqrt 3 {k_y} - 2\cos \frac{{3{k_x}}}{2}\sin \frac{{\sqrt 3 {k_y}}}{2}} \right) $ and $\sigma_{x,y,z}$ are Pauli matrices acting on the sublattice degrees of freedom. 

It is noted that the Hamiltonian preserves a unique pTRS $\Xi h(k){\Xi^{-1}} = h(- k)$ with $\Xi  = {e^{ - i {{\sigma _0}{N_{yz}}} }}\mathcal{K}$, where $\mathcal{K}$ represents the operation of complex conjugation and $\sigma_{0}$ is an identity matrix. $\sigma_{0} N_{yz}$ represents the Kronecker product of $\sigma_{0}$ and $N_{yz}$. The pTRS operator $\Xi$ is anti-unitary and satisfies $\Xi^2=-I_A$, where 
${I_A} = \left( {\begin{array}{*{20}{c}}
0&0&1\\
0&1&0\\
1&0&0
\end{array}} \right)$ is a square matrix with ones on the anti-diagonal and zeros elsewhere, and its square gives the identity matrix $I_A^2=F_0$.  It is straightforward to see that the Hamiltonian breaks TRS $\mathcal{T}=e^{-i\pi \sigma_0F_y}\mathcal{K}$ and $[ h(k), \mathcal{T}]=2\beta_k\sigma_zs_2\mathcal{K}$. Later, we will see that pTRS plays an important role in protecting the topological edge states with broken TRS.

Diagonalizing Eq.(\ref{Eq2}) yields six energy bands
\begin{eqnarray} \label{Eq3}
{E_{\mp, \pm }} =  \pm \sqrt {{{\left| \alpha  \right|}^2} + {{\left( {{\beta _k} \mp {\lambda _v}} \right)}^2}},~{E_{0, \pm }} =  \pm u,  
\end{eqnarray}
which are symmetric to $E=0$. The energy gap closes at the Dirac points $\textbf{\textit{K}} = \left( {2\pi /3,2\sqrt 3 \pi /9} \right)$ and $\textbf{\textit{K}}' =  - \textbf{\textit{K}}$. The corresponding band gap at the two high-symmetry points is $\Delta E_g^K = \Delta E_g^{K'} = \left| {3\sqrt 3 t' - {\lambda _v}} \right|$. When ${\lambda _v} > 3\sqrt 3 t'$, the gap is dominated by ${\lambda _v}$ and the system is a trivial normal insulator (NI). When $3\sqrt 3 t' > {\lambda _v}$, it becomes a topologically non-trivial QSTH insulator. The phase diagram is shown in Fig.~\ref{fig_band}(b), and two representative band spectra from QSTH and NI phases are plotted in panels (c) and (d).

\section{Edge states and topological characterization}
\begin{figure}	\includegraphics[width=1.03\columnwidth]{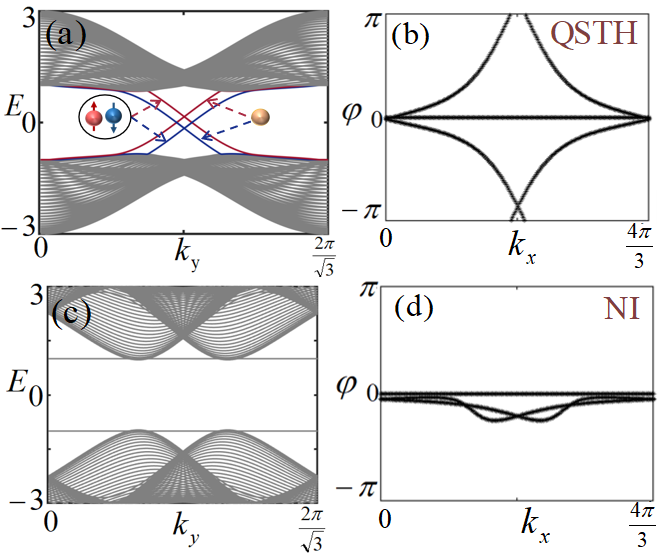} 	\caption{
             (a) and (b) The energy spectra on strips with zigzag boundary conditions and the Berry phase for $\mathbb{Z}_2$ invariant in topological QSTH phase. The red (blue) curve highlights the helical edge states at one (the other) edge, and 
            we set ${\lambda _v} = 0.2$ as in Fig.~\ref{fig_band}(c).
            (c) and (d) Similar to (a) and (b), but for trivial NI phase when ${\lambda _v} = 2$. Common parameters are $t = 1,t' = 0.2$ and $u = 1$.
        \label{fig_edge}
     }
 \end{figure}
We first compute the energy spectra on a strip with a zigzag boundary in the topological QSTH phase as shown in Fig.~\ref{fig_edge}(a). A pair of helical edge states is across the bulk gap at each edge. These edge states are robust against weak perturbations as long as the bulk energy gap is open and pTRS perseveres.

The topology of the proposed QSTH insulator can be characterized by a $\mathbb{Z}_2$ invariant from the Wilson loop \cite{w1, w2}. The Wilson line element is constructed by $
{\left[ {G\left( k \right)} \right]^{mn}} = \left\langle {{{u^m}\left( {k + \Delta k} \right)}}
 \mathrel{\left | {\vphantom {{{u^m}\left( {k + \Delta k} \right)} {{u^n}\left( k \right)}}}
 \right. \kern-\nulldelimiterspace}
 {{{u^n}\left( k \right)}} \right\rangle $,
where ${{u^n}\left( k \right)}$ is $n$-th occupied Bloch wave function with ${h}\left( k \right)\left| {{u^n}\left( k \right)} \right\rangle  = {E_n}\left( k \right)\left| {{u^n}\left( k \right)} \right\rangle $, and $\Delta k = \left( {{k_f} - {k_i}} \right)/N$ is a step defined by two momentum points ${k_{i,f}}$ and the number of unit cells $N$. A path-ordered discrete Wilson line operator is defined as ${W_{{k_i} \to {k_f}}} = G\left( {{k_f} - \Delta k} \right)G\left( {{k_f} - 2\Delta k} \right) \ldots G\left( {{k_i} + \Delta k} \right)G\left( {{k_i}} \right)$. A closed Wilson line operator ${W_{{k_i} \to {k_i} + {b_2}}}$ starts from the base momentum point ${k_i}$ and returns to ${k_f} = {k_i} + {b_2} = {k_i}$ with ${b_2} = \left( {\frac{{2\pi }}{3},\frac{{2\sqrt 3 \pi }}{3}} \right)$ a reciprocal lattice vector. The Berry phase $\varphi \left( {{k_i}} \right)$ is defined as the phase of the eigenvalues $\varepsilon \left( {{k_i}} \right) = \left| {{\varepsilon _m}} \right|{e^{i {\varphi _m}\left( {{k_i}} \right)}}$ of the Wilson line loop operator, where ${W_{{k_i} \to {k_i} + {b_2}}}\left| {{\varphi _m}\left( {{k_i}} \right)} \right\rangle  = {\varepsilon _m}\left| {{\varphi _m}\left( {{k_i}} \right)} \right\rangle $. The $\mathbb{Z}_2$ topological invariant is defined by $\nu  = \sum\nolimits_m {\left| {{Z_m}} \right|} \mod 2$, where ${Z_m} = \frac{1}{{4\pi }}\oint_l {{\nabla _k}{\varphi _m} \cdot dk}$ integrated on the closed loop $l$ including $k_i$ in the Brillouin zone and $m$ runs from 1 to ${n_{occ}=3}$ the number of the occupied Bloch bands. 

In this work, we choose a base momentum ${k_i} = \left( {{k_x},0} \right), {k_x} \in \left[ {0,4\pi /3} \right]$. We plot the evolution of the Berry phase in the QSTH phase in Fig.~\ref{fig_edge}(b). And we have three branches -- two travel in opposite directions and cross $\pm\pi$ to give a nontrivial $\nu=1$ while the middle one remains constant. 

We also compute the spectra on a ribbon and the $\mathbb{Z}_2$ invariant in the trivial NI phase, which are presented in Figs.~\ref{fig_edge}(c) and (d). No edge states are observed, and all $\varphi_m$ come to the original points without completing a full loop, leading to vanishing winding number $\nu=0$. The above result confirms the bulk-boundary correspondence in QSTH insulators.

If we take a closer inspection of the edge states, the one with positive velocity consists only of spin-0 components, while the counter-propagating one has an equal mix of spins-$\pm 1$, as in Fig.~\ref{fig_edge}(a) (see Appendix \ref{AppB} for more details). For simplicity, we consider only the spin components of the helical edges $\left|\Uparrow\right\rangle=\frac{1}{\sqrt{2}}(|1\rangle+\left|-1\right\rangle)$ and $\left|\Downarrow\right\rangle=|0\rangle$. It is easy to verify that $\left\langle\Uparrow\vert\Downarrow \right\rangle=0$, $\Xi\left|\Uparrow\right\rangle = \left|\Downarrow\right\rangle$ and $\Xi\left|\Downarrow\right\rangle = -\left|\Uparrow\right\rangle$, which is similar to how the helical edge states or the Kramers' pair transformed under TRS in QSH insulators. Thus, the edge states in QSTH insulators are protected by pTRS and are free from scattering \cite{scat1, scat2}. 

\section{Spin-tensor Hall conductivity}

With the spin components of edge states, we can evaluate different currents on the boundary. For rank-0 spin-tensor current or the charge current, it vanishes as there is a pair of counter-propagating edge states. For rank-1 spin-tensor current or the spin current, it should vanish as well since neither $\langle\psi_+|\hat{F_z}|\psi_+\rangle$ nor $\langle\psi_0|\hat{F}_z|\psi_0\rangle$ contributes to it, {where $|\psi_{+}\rangle$ and $|\psi_{0}\rangle$ denote the wave functions for the helical edge states. Finally, we can show that the rank-2 spin-tensor current associated with $N_{zz}$ is nonzero. This back-of-the-envelope calculation suggests that this is indeed a QSTH insulator with only a higher-rank spin-tensor current on the edge. In the following, we use the Kubo formula to evaluate the conductivity precisely.

Formally, the charge current, spin current, and rank-2 spin-tensor current operator can be defined as
\begin{equation}
\hat{J_0} = q{F_0} \hat v,~\hat{J_1^z} = \frac{1}{2}{\left\{ {\hbar {F_z},\hat v} \right\}_ + },~\hat{J_2^{zz}} = \frac{1}{2}{\left\{ {\hbar {N_{zz}},\hat v} \right\}_ + }
\label{Eq4}
\end{equation}%
where $F_0$ is the identity matrix, ${F_z} = diag\left( {1,0, - 1} \right)$ and ${N_{zz}} = diag\left( {\frac{1}{3}, - \frac{2}{3},\frac{1}{3}} \right)$. Here, $\hat{v} = {\partial _p} h\left( k \right)$ is the velocity operator. The rank-2 spin-tensor Hall conductivity can be calculated using the Kubo formula at the clean limit
\begin{eqnarray}
\notag
\sigma _{xy}^{zz} &=& \frac{\hbar }{V}\sum\limits_{n \ne n',k} {\left( {{f_{n'k}} - {f_{nk}}} \right)}\notag\\
&\times& \frac{{{\mathop{\rm Im}\nolimits} \left[ {\left\langle {n'k} \right|{\hat{J}_{2}^{zz}}\left| {nk} \right\rangle \left\langle {nk} \right|{\hat{J}_{0,y}}\left| {n'k} \right\rangle } \right]}}{{{{\left( {{E_{nk}} - {E_{n'k}}} \right)}^2}}}
\label{Eq5}
\end{eqnarray}%
where $n\left( {n'} \right)$ is the band index, and ${f_{nk}} = {\left[ {{e^{\left( {{E_{mk}} - {E_F}} \right)/{k_B}T}} + 1} \right]^{ - 1}}$ is the Fermi distribution given the Fermi energy $E_F$, and $V$ is the volume of the unit cell. For charge and spin currents, $\sigma _{xy}$ and $\sigma _{xy}^{z}$ can be computed similarly by replacing the current operator correspondingly.

The results for $\sigma _{xy}$ (HC), $\sigma _{xy}^{z}$ (SHC) and $\sigma _{xy}^{zz}$ (STHC) are plotted in Fig.~\ref{fig_Kubo}(a). It is obvious that all the conductivity vanishes in the trivial insulator phase, and only the rank-2 spin-tensor Hall conductivity is quantized to a constant that is independent of details of the systems like coupling strengths:
\begin{equation}
    \sigma _{xy}=0,~\sigma _{xy}^{z} =0~\text{and}~\sigma _{xy}^{zz}=\frac{q}{4\pi}.
    \label{eq6}
\end{equation}
Besides the rank-2 spin tensor current $J_2^{zz}$ of particular interest here, we can define other spin-tensor currents like $J_2^{xy}$ or $J_2^{yz}$, and the corresponding conductivities are verified to be zero. However, we want to highlight the constraint as
\begin{equation}
    \sigma _{xy}^{xx} + \sigma _{xy}^{yy} + \sigma _{xy}^{zz} = 0
\end{equation}
because $\sum\nolimits_i {{N_{ii}} = 0} \left( {i = x,y,z} \right)$. Applying the Kubo formula, we find $\sigma _{xy}^{xx} = 0$ and $\sigma _{xy}^{yy} =  - \sigma _{xy}^{zz}$ so that the equation holds.

\begin{figure} \includegraphics[width=0.98\columnwidth]{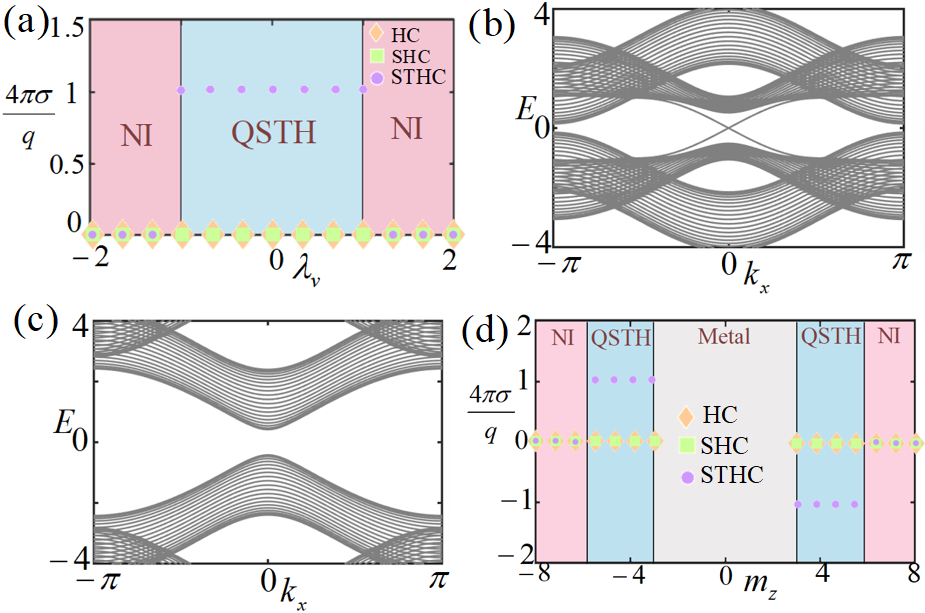}
    \caption{
       (a) Hall conductivity at different ranks computed from the bulk using the Kubo formula for the honeycomb lattice. All those, including Hall conductivity (HC) for rank-0 charge current, spin-Hall conductivity (SHC) for rank-1 spin current, and spin-tensor-Hall conductivity (STHC) for rank-2 spin-tensor current, are plotted across both NI and QSTH phases. Only in the nontrivial QSTH phase is a quantized STHC observed. (b) Energy spectra for a strip in QSTH phase on a square lattice under periodic boundary conditions along $x$ but open boundary conditions along $y$. The parameter $m_z$ is set to be $m_z=3.2$. (c) Similar to (b) but for the model on a square lattice in NI phase with $m_z=7.2$. (d) Similar to (a) but for the model on a square lattice. 
     }
     \label{fig_Kubo}
 \end{figure}

\section{A toy model for QTSH on a sqaure lattice}

In addition to the model on the honeycomb lattice, we present another toy model on a square lattice. Its Hamiltonian under the basis $\hat C_k^\dagger  = \left( {\hat a_{k,1}^\dagger ,\hat b_{k,1}^\dagger ,\hat a_{k,0}^\dagger ,\hat b_{k,0}^\dagger ,\hat a_{k, - 1}^\dagger ,\hat b_{k, - 1}^\dagger } \right)$ in momentum space reads
\begin{equation}
    h_{squ}(k) = \Gamma _{25}\sin {k_x} +  \Gamma _{26}\sin k_y + \Gamma _2 \xi_k - \Gamma _4 m_z,
    \label{eq8}
\end{equation}
where $\xi_k=\cos k_x + \cos k_y$, ${\Gamma _{(1,2,3,4,5,6)}} = ({\sigma _x} \otimes {F_x},{\sigma _z} \otimes {F_x},{\sigma _x} \otimes {N_{yz}},{\sigma _z} \otimes {N_{yz}},{\sigma _y} \otimes {N_{xx}},{\sigma _y} \otimes {F_0})$ and ${\Gamma _{\alpha \beta }} = \left[ {{\Gamma _\alpha },{\Gamma _\beta }} \right]/2i$ are commutators. This model also preserves pTRS while breaking TRS. We choose the hopping strength to be units for the energy, and $m_z$ represents a constant external field that could drive the topological phase transition from NI to QSTH. See Appendix \ref{AppC} for the corresponding tight-binding Hamiltonian $\hat H_{squ}$ on the square lattice. 

To see the edge states, we consider a strip of the two-dimensional insulator. We take periodic boundary conditions along $x$, but open boundary conditions along $y$. Since the translation invariance holds along $x$, we could partially Fourier transform the Hamiltonian along $x$. After the Fourier transformation, the original Hamiltonian is composed of a set of one dimensional lattice Hamiltonians indexed by a continuous parameter $k_x$, namely, $\hat H_{squ}=\sum_{k_x}\hat H'_{squ}(k_x)$, where the $k_x$-dependent Hamiltonian reads
\begin{eqnarray}
\hat H'_{squ}\left( {{k_x}} \right) & =&  \sum\limits_{{i_y} = 1}^{{N_y} - 1} {\left( {\hat C_{{i_y} + 1,{k_x}}^\dag \frac{{{\Gamma _2} + i{\Gamma _{26}}}}{2}{{\hat C}_{{i_y},{k_x}}} + h.c.} \right)} \notag\\
 & +&  \sum\limits_{{i_y} = 1}^{{N_y}} {\left[ {\hat C_{{i_y},{k_x}}^\dag \left( {\sin {k_x} \cdot {\Gamma _{25}} + \cos {k_x} \cdot {\Gamma _2}} \right){{\hat C}_{{i_y},{k_x}}}} \right]} \notag\\
 & -&  \sum\limits_{{i_y}= 1}^{{N_y}} {\left( {\hat C_{{i_y},{k_x}}^\dag {m_z}{\Gamma _4}{{\hat C}_{{i_y},{k_x}}}} \right)}. 
\end{eqnarray}
We compute the edge states that exhibit similar behaviors in the QSTH and NI phase as shown in Figs.~\ref{fig_Kubo}(b) and (c). The behavior of the edge states signals a clear difference between the two phases. In the QSTH phase, there is an edge state across the bulk gap at each edge. However, in the NI phase, there are no edge states. In the QSTH phase, a quantized non-zero spin tensor Hall conductivity is also present, as in Fig.~\ref{fig_Kubo}(d). We also find that its topology can be characterized by the spin-Chern number \cite{sc1,sc2,sc3} (see Appendix \ref{AppC} for more details). This suggests that QSTH insulators can be versatile and opens the possibility of exploring them in different physical systems.

 \section{Physical interpretation and connections to QSH insulators}
 
While the models for QSTH insulators are intrinsically complicated, we identify intuitive and physical interpretations of such an exotic state of matter by revealing its underlying connections to the QSH insulators.

If we cast the model Hamiltonian in Eq.~(\ref{Eq2}) onto a new basis ${\hat c}_{k}={\left( {{\hat a}_{k, \Uparrow},{\hat b}_{k, \Uparrow},{\hat a}_{k, \Downarrow }, {\hat b}_{k, \Downarrow }} \right)^T}$ that is defined by $\left|\Uparrow\right\rangle=\frac{1}{\sqrt{2}}(|1\rangle+\left|-1\right\rangle)$ and $\left|\Downarrow\right\rangle=|0\rangle$, we would arrive at the Kane-Mele model as
\begin{equation}
    H_{KM} = {\mathop{\rm Re}\nolimits} \left( {{\alpha _k}} \right) \cdot {\sigma _x}{\tau_0} + {\mathop{\rm Im}\nolimits} \left( {{\alpha _k}} \right) \cdot {\sigma _y}{\tau_0} + {\beta _k} \cdot {\sigma _z}{\tau_z} + {\lambda _v} \cdot {\sigma _z}{\tau_0}
\end{equation}
when $u=0$. As a result, the edge states of QSTH insulators can be projected similarly to those of QSH insulators. This gives the physical origins of the topology of the proposed QSTH insulators and how they can be characterized by a $\mathbb{Z}_2$ invariant.

However, this does not suggest that the coupling term of $u$ is trivial. In contrast, it plays a vital role in shaping the overall band structure and driving the topological phase. To see this, we need to rewrite the corresponding term as $u\left| {\o} \right\rangle \left\langle {\o} \right|$, where $\left| {\o} \right\rangle  = \frac{1}{{\sqrt 2 }}\left( {\left| 1 \right\rangle  - \left| { - 1} \right\rangle } \right)$. First, we notice that $\left|  \Uparrow  \right\rangle$ and $\left|  \Downarrow  \right\rangle$ are orthogonal to $\left|\o  \right\rangle$. That being said, $\left| {\o} \right\rangle$ is a dark state, and the system would always remain gapless when $u=0$ so that it could never host any gaped topological states. More importantly, this term does not affect the system's topological characterizations since the dark state itself is invariant under pTRS $\Xi|\o\rangle = |\o\rangle$.

Mathematically, it indicates that the proposed QSTH insulators can be characterized by a subgroup SU(2)$\times$U(1), in which the matrix representation of pTRS reads
\begin{eqnarray}
    \Xi = \begin{pmatrix}
        i\sigma_y & 0 \\
        0 & 1 \\
    \end{pmatrix} \mathcal{K}
\end{eqnarray}
in the basis $\{\left|  \Uparrow  \right\rangle, \left|  \Downarrow  \right\rangle, \left| {\o} \right\rangle\}$. Such an observation confirms that the nontrivial QSTH phase is indeed protected by pTRS and resonates with the unique pattern of the Berry phase of the $\mathbb{Z}_2$ invariant in Fig.~\ref{fig_edge}(d).

Next, we would like to show that the rank-2 spin tensor current is actually equivalent to the rank-1 pseudo-spin current. In the new basis, the pseudo spins $\left|  \Uparrow  \right\rangle  $ and $\left|  \Downarrow  \right\rangle$ are orthogonal to $\left| {\o} \right\rangle$. It is noted that, when projecting into the pseudo-spin subspace, the spin tensor $N_{zz}$ becomes
\begin{eqnarray}
{P_s}{N_{zz}}P_s^{ - 1} = \frac{1}{2}{\sigma _z} - \frac{1}{6}{\sigma _0},
\end{eqnarray}
where ${P_s} = \left|  \Uparrow  \right\rangle \left\langle  \Uparrow  \right| + \left|  \Downarrow  \right\rangle \left\langle  \Downarrow  \right|$ is the projection operator. The velocity operator $ \hat v = {\partial _p} h\left( k \right)$ projecting into the pseudo-spin subspace becomes ${P_s}\hat vP_s^{ - 1} = \tilde {\hat v}$, where $\tilde {\hat v} = {\partial _p} H_{KM}$. The rank-2 spin tensor current operator then becomes
\begin{eqnarray}
\hat J_2^{zz} = \frac{1}{2}{\left\{ {\hbar \frac{{{\sigma _z}}}{2},\tilde {\hat v}} \right\}_ + } - \frac{1}{6}{\sigma _0}\tilde{\hat v}.
\label{11}
\end{eqnarray}

Here, the first term is just the rank-1 pseudo-spin current operator while the second term is proportional to the charge current operator. Due to the pTRS, the second term in Eq. (\ref{11}) has no contribution to the rank-2 spin tensor current. Therefore, the rank-2 spin tensor current in the projected subspace is equivalent to the rank-1 pseudo-spin current. The rank-2 spin tensor Hall conductivity is
\begin{eqnarray}
\sigma _{xy}^{zz} &=& \tilde \sigma _{xy}^z = \frac{\hbar }{{4q}}\left( {\tilde \sigma _{xy}^ \Uparrow  - \tilde \sigma _{xy}^ \Downarrow } \right) 
\notag\\
&=& \frac{\hbar }{{4q}}\left[ {\frac{{{q^2}}}{h} - \left( { - \frac{{{q^2}}}{h}} \right)} \right] = \frac{q}{{4\pi }},
\end{eqnarray}
 where $\tilde \sigma _{xy}^z$  is the rank-1 quantum pseudospin  Hall conductivity. $\tilde \sigma _{xy}^\Uparrow $ and $ \tilde \sigma _{xy}^ \Downarrow$ denote the quantum Hall conductivity of particles with $\Uparrow$ and $\Downarrow$, respectively. Therefore, the result in the projected pseudo-spin subspace is consistent with that derived from the Kubo formula as in Eq. (\ref{eq6}).

Following the same arguments, the QSTH model on a square lattice can be mapped to a spin-$\frac{1}{2}$ QSH model under the
basis ${\hat c}_{k}$ as $H_{S} = -{\sin {k_x}\cdot\sigma _x}{\tau_x} -\frac{1}{2} \sin {k_y}\cdot {\sigma _x}{\tau_y} + M{(k)}\cdot {\sigma _z}{\tau_0}$, with $M{(k)}=(\cos {k_y}+ \cos {k_x}-\frac{1}{3}m_z)$. An additional term needs to be included $(\cos {k_y}+ \cos {k_x}+m_z)\left| {\o} \right\rangle \left\langle {\o} \right|$. Taking a similar procedure, we also obtain the rank-2 QSTH conductivity in the projected pseudo-spin subspace same as that by the Kubo formula.

These key observations bridge QSTH insulators and QSH insulators from a physical perspective, similar to the way that QSH insulators were introduced by doubling the Chern insulators.

 \section{Discussion}
 
This work establishes the theory for QSTH insulators and enriches the family of Hall effects. We start with the Hamiltonian on a honeycomb lattice to realize QSTH insulators, a new type of topological state protected by pTRS. We provide a full characterization of the topological properties of QSTH insulators and identify the rank-2 spin-tensor Hall conductivity as a universal constant that is independent of detailed model parameters, while both rank-0 charge and rank-1 spin Hall conductivities vanish. We further provide another toy model on a square lattice and give a physical interpretation of the QSTH insulator by bridging it with the QSH insulator.

In terms of the experimental realization of the proposed systems. The ultracold atoms offer a highly tunable and controllable platform to realize many members in the family of Hall effects, including AHE, QAHE, SHE, and QSHE \cite{4,QWZ,cc1,cc2,cc3,cc4,cc5}. Several key components like pesudospin-1 ultracold atomic systems \cite{ll1,ll2,ll3,ll4,ll5} and honeycomb lattice \cite{h1,Bloch16,Schneider16,h3} have been experimentally demonstrated. The key to driving QSTH insulators is the spin-tensor-momentum coupling whose experimental proposals are discussed in \cite{stm1,stm2}.

Last but not least, there remain tons of physics questions to be answered for QSTH insulators, such as whether there are other types of rank-2 QSTH insulators in spin-1 systems, the general construction of rank-$n$ QSTH insulators in arbitrarily spinful systems, how QSTH phases respond to interactions and disorders, etc. Our work adds a new member to the celebrated Hall effect family as well as the exciting world of topological states of matter. Moreover, it provides new insights into physics raised by spin tensors in large-spin systems to enable futuristic functionalities in spintronics and atomtronics.
\begin{acknowledgments}
J. Hou thanks C. Zhang and Y. Su for inspiring discussions. Y. Wu and T. Li and are supported by NSFC under grant No.12275203, Innovation Capability Support Program of Shaanxi (2022KJXX-42), and 2022 Shaanxi University Youth Innovation Team Project (K20220186).
\end{acknowledgments}

\appendix

\section{Definition of spin operators in a spin-1 system}
\setcounter{equation}{0} 
\renewcommand{\theequation}{A\arabic{equation}} 
\setcounter{figure}{0} 
\renewcommand{\thefigure}{A\arabic{figure}}
\label{AppA}
Under the basis $\left\{ {\left| 1 \right\rangle ,\left| 0 \right\rangle ,\left| { - 1} \right\rangle } \right\}$, the 8 Gell-Mann matrices that constitute the generators of the SU(3) group are
are defined as
\begin{eqnarray}
\lambda _{1} &=&%
\begin{pmatrix}
0 & 1 & 0 \\
1 & 0 & 0 \\
0 & 0 & 0%
\end{pmatrix}%
,\lambda _{2}=%
\begin{pmatrix}
0 & -i & 0 \\
i & 0 & 0 \\
0 & 0 & 0%
\end{pmatrix}%
,\lambda _{3}=%
\begin{pmatrix}
1 & 0 & 0 \\
0 & -1 & 0 \\
0 & 0 & 0%
\end{pmatrix}%
,  \notag \\
\lambda _{4} &=&%
\begin{pmatrix}
0 & 0 & 1 \\
0 & 0 & 0 \\
1 & 0 & 0%
\end{pmatrix}%
,\lambda _{5}=%
\begin{pmatrix}
0 & 0 & -i \\
0 & 0 & 0 \\
i & 0 & 0%
\end{pmatrix}%
,\lambda _{6}=%
\begin{pmatrix}
0 & 0 & 0 \\
0 & 0 & 1 \\
0 & 1 & 0%
\end{pmatrix}%
,  \notag \\
\lambda _{7} &=&%
\begin{pmatrix}
0 & 0 & 0 \\
0 & 0 & -i \\
0 & i & 0%
\end{pmatrix}%
,\lambda _{8}=\frac{1}{\sqrt{3}}%
\begin{pmatrix}
1 & 0 & 0 \\
0 & 1 & 0 \\
0 & 0 & -2%
\end{pmatrix}%
.
\end{eqnarray}

The spin vectors can be expanded by the Gell-Mann matrices as
\begin{eqnarray}
{F_x} &=& \frac{{{\lambda _1}}}{{\sqrt 2 }} + \frac{{{\lambda _6}}}{{\sqrt 2 }} = \frac{1}{{\sqrt 2 }}\left( {\begin{array}{*{20}{c}}
0&1&0\\
1&0&1\\
0&1&0
\end{array}} \right),
\notag \\
{F_y} &=& \frac{{{\lambda _2}}}{{\sqrt 2 }} + \frac{{{\lambda _7}}}{{\sqrt 2 }} = \frac{1}{{\sqrt 2 }}\left( {\begin{array}{*{20}{c}}
0&{ - i}&0\\
i&0&{ - i}\\
0&i&0
\end{array}} \right),
\notag \\
{F_z} &=& \frac{{{\lambda _3}}}{2} + \frac{{\sqrt 3 {\lambda _8}}}{2} = \left( {\begin{array}{*{20}{c}}
1&0&0\\
0&0&0\\
0&0&{ - 1}
\end{array}} \right),
\end{eqnarray}
and the spin tensors are defined by the anticommutator
\begin{eqnarray}
{N_{ij}} = \frac{1}{2}{\left\{ {{F_i},{F_j}} \right\}_ + } - {\delta _{ij}}\frac{{{{\bf{F}}^2}}}{3}
\end{eqnarray}

Because ${N_{ij}} = {N_{ji}}$ and $\sum\nolimits_i {{N_{ii}}}  = 0$ according to the definition, only five of the nine spin tensors are linearly independent and form another set of generators of the SU(3) group. The five linearly independent spin tensors are

\begin{eqnarray}
N _{xx} &=&%
\left( {\begin{array}{*{20}{c}}
{ - \frac{1}{6}}&0&{\frac{1}{2}}\\
0&{\frac{1}{3}}&0\\
{\frac{1}{2}}&0&{ - \frac{1}{6}}
\end{array}} \right)
,N _{yy}=%
\left( {\begin{array}{*{20}{c}}
{ - \frac{1}{6}}&0&{ - \frac{1}{2}}\\
0&{\frac{1}{3}}&0\\
{ - \frac{1}{2}}&0&{ - \frac{1}{6}}
\end{array}} \right),
 \notag \\
N _{zz}&=&%
\left( {\begin{array}{*{20}{c}}
{\frac{1}{3}}&0&0\\
0&{ - \frac{2}{3}}&0\\
0&0&{\frac{1}{3}}
\end{array}} \right)
, 
N _{xy} =%
\frac{1}{2}\left( {\begin{array}{*{20}{c}}
0&0&{ - i}\\
0&0&0\\
i&0& 0 
\end{array}} \right),
 \notag \\
N _{xz}&=&%
\frac{1}{{2\sqrt 2 }}\left( {\begin{array}{*{20}{c}}
0&1&0\\
1&0&{ - 1}\\
0&{ - 1}&0
\end{array}} \right)
.
\end{eqnarray}

In the main text, $s_1 = {\bf{F}}^2 - 2F_z^2 + s_2 $, $s_2 = -N_{yy} + N_{zz}$, and $s_3 = \frac{1}{2}(-{\bf{F}}^2+4F_z^2) - s_2 $. Their explicit matrix forms  are given by

\begin{eqnarray}
s _{1} &=&%
\left( {\begin{array}{*{20}{c}}
{\frac{1}{2}}&0&{\frac{1}{2}}\\
0&1&0\\
{\frac{1}{2}}&0&{\frac{1}{2}}
\end{array}} \right)
,s _{2}=%
\left( {\begin{array}{*{20}{c}}
{\frac{1}{2}}&0&{\frac{1}{2}}\\
0&{ - 1}&0\\
{\frac{1}{2}}&0&{\frac{1}{2}}
\end{array}} \right),
\notag \\
s _{3}&=&%
\left( {\begin{array}{*{20}{c}}
{\frac{1}{2}}&0&{ - \frac{1}{2}}\\
0&0&0\\
{ - \frac{1}{2}}&0&{\frac{1}{2}}
\end{array}} \right)
.  \notag \\
\end{eqnarray}

\section{Details of the spin components}
\label{AppB}
\setcounter{equation}{0} 
\renewcommand{\theequation}{B\arabic{equation}} 
\setcounter{figure}{0} 
\renewcommand{\thefigure}{B\arabic{figure}}
In the quantum spin tensor Hall (QSTH) phase, we conduct a more detailed analysis of the gapless boundary states and their spin components. There are four gapless boundary states in total, with two having positive velocity and two
having negative velocity. On each side of the system, there is one positive and one negative velocity state. Analyzing
one side, the one with positive velocity consists only of spin-0 components, while the counter-propagating one has an equal mix of spins-$ \pm 1$, as shown in Fig.~\ref{Fig01}. For simplicity, we consider only the spin components of the helical edge states on one edge, i.e., $|\psi_{v,+}\rangle=\frac{1}{\sqrt{2}}(|1\rangle+|-1\rangle)$ and $|\psi_{-v,0}\rangle=|0\rangle$. It's easy to verify that $\langle\psi_{v,+}\vert\psi_{-v,0} \rangle=0$, $\Xi|\psi_{v,+}\rangle = |\psi_{-v,0}\rangle$ and $\Xi|\psi_{-v,0}\rangle = -|\psi_{v,+}\rangle$, which is similar to how are the helical edge states transformed under TRS in QSH. Thus, the edge states in QSTH are protected by pTRS and are free from scattering.

Thus, in the QSTH phase, charge currents in opposite directions cancel each other out, resulting in a net charge flow of zero for the boundary states, while spin currents in the same direction cancel each other out, leading to a net spin flow of zero; however, the spin tensor flow is not zero, arising from the topologically protected dissipationless edge states.
\begin{figure}	\includegraphics[width=0.96\columnwidth]{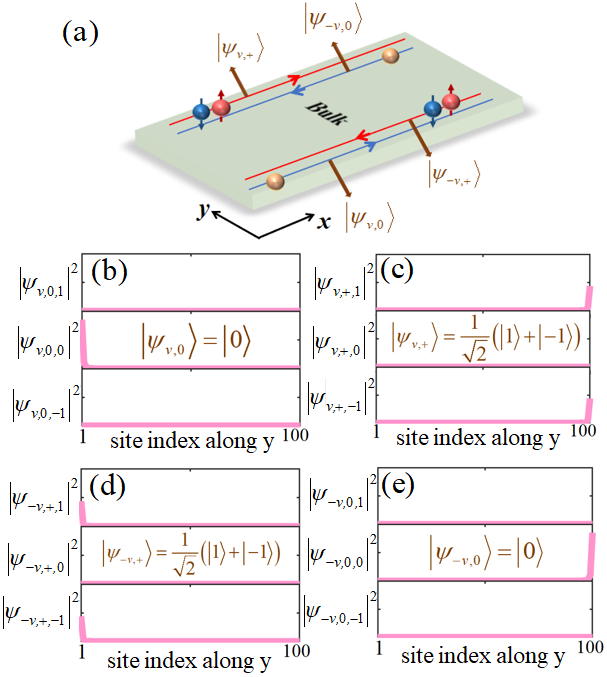}
 	\caption{(a) Illustration of edge states on a stripe in QSTH phase.
    (b) and (d) The density distribution of the helical edge states on one edge. (c) and (e) The density distribution of the helical edge states on the other edge. ${\left| {{\psi _{\pm v,0,s}}} \right|^2}$ and ${\left| {{\psi _{\pm v,+,s}}} \right|^2}$ denote the particle density for different spin components (${s = 1,0, - 1}$) of the edge states. The number of lattice sites along $y$ is $N_y=100$.} Parameters are $t = 1,t' = 0.2$, ${\lambda _v} = 0.2t$ and $u = 1$.
      \label{Fig01}
 \end{figure}
 
 \section{ Calculation of the spin Chern number for QSTH on a square lattice}
 \label{AppC}
 \setcounter{equation}{0} 
\renewcommand{\theequation}{C\arabic{equation}} 
\setcounter{figure}{0} 
\renewcommand{\thefigure}{C\arabic{figure}}
\begin{figure}	\includegraphics[width=0.9\columnwidth]{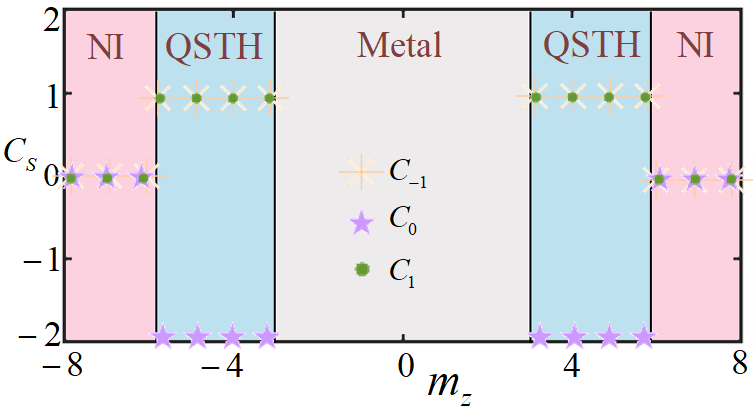} \caption{
Spin Chern number ${C_s}$, ($S=\pm 1,0$) at different ranks computed from the bulk for the square lattice. All those are plotted across both NI and QSTH phases. Only in the nontrivial QSTH phase is no zeros. }
     \label{Fig02}
 \end{figure} 
In the main text, we construct a toy model on a square lattice as in Eq. (\ref{eq8}) to realize the QSTH effect. We obtain its tight-binding Hamiltonian on the lattice by the inverse Fourier transformation as
\begin{eqnarray}
{\hat{H}_{squ}}  &=&  \sum\limits_{{i}}{{\left( {\hat c_{{i}+e_x}^\dag {{\tilde{\Gamma} _x}}{{\hat c}_{{i}}}+\hat c_{{i} +e_y}^\dag {{\tilde{\Gamma} _y}}{{\hat c}_{{i}}} + h.c.}\right)}}  \notag\\
  &-&  \sum\limits_{{i}} {{{\hat c_{{i}}^\dag {m_z}{\Gamma _4}{{\hat c}_{{i}}}}}}, 
\end{eqnarray}
where $c_{i}^\dagger =(c_{i,-1}^\dagger ~ c_{i,0}^\dagger ~ c_{i,1}^\dagger)$, $\tilde{\Gamma} _x=(\Gamma _2+i{\Gamma _{25}})/2$, and $\tilde{\Gamma} _y=(\Gamma _2+i{\Gamma _{26}})/2$. $e_x$ and $e_y$ denote the unit vectors along $x$ and $y$, respectively. The model describes a particle with pseudo spin (three internal states) that hops on a lattice where the nearest neighbor hopping is accompanied by an operation on the pseudo-spin degrees of freedom. The operation on pseudo-spin degrees of freedom is different for the hoppings along the $x$ and $y$ directions. In addition, there is a staggered onsite potential with the strength $m_z$.

Next, we consider a generalization of the spin Chern number to characterize the topological phases of QSTH insulators. First, we construct a matrix $M(k) = \left\langle {{u_n}(k)} \right|{\sigma _0} \otimes {\lambda _2}\left| {{u_n}(k)} \right\rangle $ whose diagonalization decomposes the mixed occupied bands into two spin sectors (denoted by $S =  \pm 1$ and $0$) satisfying $M(k)\left| {{\psi _S}(k)} \right\rangle  = {w_S}\left| {{\psi _S}(k)} \right\rangle $.
When the eigenspectra ${w_S}$ of three spin sectors are separable, we can define the spin Chern number for each spin sector ${C_S} = \frac{1}{{2\pi }}\smallint {d^2}k \cdot {F_S}(k)$ through the Berry curvature ${F_S}(k) = \nabla \times {A_S}(k)$, where the non-Abelian Berry connection
\begin{eqnarray}
 {A_S}(k) =  - i\left\langle {{\psi _S}(k) \cdot u(k)} \right|{\partial _k}\left| {{\psi _S}(k) \cdot u(k)} \right\rangle  
\end{eqnarray}
 and $\left| {{\psi _S}(k) \cdot u(k)} \right\rangle  = \sum\nolimits_j {\left| {{\psi _j}(k) \cdot {u^j}(k)} \right\rangle }$. The summation of $j$ runs over all occupied bands and  ${\psi _j}\left( k \right)$ denotes the $j$th component of the eigenvector $\left| {{\psi _S}\left( k \right)} \right\rangle $. In our context, a nonzero spin Chern number $C_S$ means there is a chiral edge state of “spin-$S$" with the chirality determined by the sign of $C_S$. In the trivial insulator phase ${C_{ \pm 1,0}} = 0$, while ${C_{ \pm 1}} = 1$ and ${C_0} =  - 2$ in QSTH phase as shown in Fig. \ref{Fig02}.


\begin{thebibliography}{99}

\bibitem{1} {E. H. Hall, On the ``Rotational coefficient" in nickel and cobalt, } \href{https://iopscience.iop.org/article/10.1088/1478-7814/4/1/335}%
{Philosophical Magazine Series 5 12, 157172 (1881).}

\bibitem{2} {T. Jungwirth, Q. Niu, and A. H. MacDonald, Anomalous Hall effect in ferromagnetic semiconductors,} \href{https://journals.aps.org/prl/abstract/10.1103/PhysRevLett.88.207208}%
{Phys. Rev. Lett. 88, 207208 (2002).}

\bibitem{3} {F. D. M. Haldane, Model for a quantum Hall effect without landau levels: Condensed-matter realization of the ``Parity anomaly", } \href{https://journals.aps.org/prl/abstract/10.1103/PhysRevLett.61.2015}%
{Phys. Rev. Lett. 61, 2015,2018 (1988).}
\bibitem{444} {C. Z. Chang, J. Zhang, X. Feng, and et.al, Experimental Observation of the Quantum Anomalous Hall Effect in a Magnetic Topological Insulator, } \href{https://www.science.org/doi/10.1126/science.1234414}%
{Science 340, 167 (2013).}




\bibitem{5} {M. I. Dyakonov and V. I. Perel, Current-induced spin orientation of electrons in semiconductors, } \href{https://www.sciencedirect.com/science/article/abs/pii/0375960171901964}%
{Phys. Lett. A 35, 459 (1971).}

\bibitem{6} {J. Sinova, S. O. Valenzuela, J. Wunderlich, C. H. Back, and T.
Jungwirth, Spin Hall effects,} \href{https://doi.org/10.1103/RevModPhys.87.1213}%
{Rev. Mod. Phys. 87, 1213 (2015).}

\bibitem{7} {S. O. Valenzuela, and M. Tinkham, Direct electronic measurement
of the spin Hall effect,} \href{https://sc.panda985.com/scholar?hl=zh-CN&as_sdt=0%2C5&q=S.+O.+Valenzuela+and+M.+Tinkham%2C+Nature+442%2C+176+%282006%29&btnG=}%
{Nature 442, 176 (2006).}

\bibitem{8} {S. Zhang, Spin Hall Effect in the Presence of Spin Diffusion,}
\href{https://doi.org/10.1103/PhysRevLett.85.393}{Phys. Rev. Lett. 85, 393
(2000).}

\bibitem{10} {C. L. Kane and E. J. Mele, Quantum Spin Hall Effect in Graphene,} \href{https://doi.org/10.1103/PhysRevLett.95.226801}{Phys. Rev.
Lett. 95, 226801 (2005).}

\bibitem{11} {C. L. Kane and E. J. Mele, $Z_2$ topological order and the quantum spin Hall effect,} \href{https://journals.aps.org/prl/abstract/10.1103/PhysRevLett.95.146802}{Phys. Rev. Lett. 95, 146802 (2005).}

\bibitem{9} {B. A. Bernevig and S.-C. Zhang, Quantum Spin Hall Effect,}
\href{https://doi.org/10.1103/PhysRevLett.96.106802}{Phys. Rev. Lett. 96,
106802 (2006).}

\bibitem{12a} {M. Z. Hasan and  C. L. Kane, Topological Insulators,} \href{https://doi.org/10.1103/RevModPhys.82.3045}{Rev. Mod. Phys. 82, 3045 (2010).}

\bibitem{12} {X. L. Qi and S.-C. Zhang, Topological Insulators and Superconductors,} \href{https://journals.aps.org/rmp/abstract/10.1103/RevModPhys.83.1057}{Rev. Mod. Phys. 83, 1057 (2011).}

\bibitem{13} {B. A. Bernevig, T. L. Hughes, and S.-C. Zhang, Quantum Spin
Hall Effect and Topological Phase Transition in HgTe Quantum Wells,} \href{https://doi.org/10.1126/science.1133734}%
{Science 314, 1757 (2006).}

\bibitem{14} {M. Konig, S. Wiedmann, C. Brune, A. Roth, H. Buhmann, L. W.
Molenkamp, X.-L. Qi, and S.-C. Zhang, Quantum Spin Hall Insulator State in
HgTe Quantum Wells,} \href{https://www.science.org/doi/10.1126/science.1148047}%
{Science 318, 766 (2007).}

\bibitem{15} {Y. K. Kato, R. C. Myers, A. C. Gossard, D. D. Awschalom,
Observation of the spin hall effect in semiconductors,} \href{https://www.science.org/doi/10.1126/science.1105514}%
{Science 306, 1910 (2004).}

\bibitem{16} {R. Raimondi, and P. Schwab, Spin-Hall effect in a disordered
2D electron-system,} \href{https://doi.org/10.1103/PhysRevB.71.033311}{Phys.
Rev. B 71, 033311 (2005).}

\bibitem{17} {S. A. Wolf, D. D. Awschalom, R. A. Buhrman, J. M. Daughton, S. von Molnár, M. L. Roukes, A. Y. Chtchelkanova, and D. M. Treger, Spintronics: A Spin-Based Electronics Vision for the Future,} \href{https://www.science.org/doi/10.1126/science.1065389}{Science 294, 1488 (2001).}

\bibitem{18} {J. Sinova, D. Culcer, Q. Niu, N. A. Sinitsyn, T. Jungwirth, and A. H. MacDonald, Universal Intrinsic Spin Hall Effect,} \href{https://journals.aps.org/prl/abstract/10.1103/PhysRevLett.92.126603}{Phys. Rev. Lett. 92, 126603 (2004).}

\bibitem{19} {L. Liu, C.-F. Pai, Y. Li, H. W. Tseng, D. C. Ralph, and R. A. Buhrman, Spin-Torque Switching with the Giant Spin Hall Effect of Tantalum,} \href{https://www.science.org/doi/10.1126/science.1218197}{Science 336, 555 (2012).}

\bibitem{20} {J. Han, A. Richardella, S. A. Siddiqui, J. Finley, N. Samarth, and L. Liu, Room-Temperature Spin-Orbit Torque Switching Induced by a Topological Insulator,} \href{https://journals.aps.org/prl/abstract/10.1103/PhysRevLett.119.077702}{Phys. Rev. Lett. 119, 077702 (2017).}

\bibitem{PrivmanQuantum1998} {V. Privman, I.D. Vagner and G. Kventsel, Quantum computation in quantum-Hall systems,} \href{https://doi.org/10.1016/S0375-9601(97)00974-2}{Phys. Lett. A 239, 141 (1998).}

\bibitem{LuFractional2024} {Z. Lu, et al., Fractional quantum anomalous Hall effect in multilayer graphene,} \href{https://doi.org/10.1038/s41586-023-07010-7}{Nature 626, 759 (2024).}

\bibitem{c2} {H. M. Price, O. Zilberberg, T. Ozawa, I. Carusotto, and N. Goldman, Four-dimensional quantum Hall effect with ultracold atoms, } \href{https://journals.aps.org/prl/abstract/10.1103/PhysRevLett.115.195303}{Phys. Rev. Lett. 115, 195303 (2015).}

\bibitem{c3} {M. Lewenstein, A. Sanpera , V. Ahufinger, et al, Ultracold atomic gases in optical lattices: mimicking condensed matter physics and beyond,} \href{https://www.tandfonline.com/doi/full/10.1080/00018730701223200}{Adv.
Phys. 56, 243 (2007).}

\bibitem{c4} {I. Bloch, J. Dalibard, S Nascimbène, Quantum simulations with ultracold quantum gases,} \href{https://www.nature.com/articles/nphys2259}{Nature Phys. 8, 267 (2012).}

\bibitem{c5} {C Gross, I. Bloch, Quantum simulations with ultracold atoms in optical lattices,} \href{https://www.science.org/doi/10.1126/science.aal3837}{Science 357, 995 (2017).}

\bibitem{c6} {D. W. Zhang, Y. Q. Zhu, Y. X. Zhao, et al, Topological quantum matter with cold atoms,} \href{https://www.tandfonline.com/doi/full/10.1080/00018732.2019.1594094}{Adv. Phys. 67, 253 (2018).}

\bibitem{l1} {G. I. Martone, F. V. Pepe, P. Facchi, S. Pascazio, and S. Stringari, Tricriticalities and Quantum Phases in Spin-Orbit-Coupled Spin-1 Bose Gases,} \href{https://journals.aps.org/prl/abstract/10.1103/PhysRevLett.117.125301}{%
Phys. Rev. Lett. 117, 125301 (2016).}

\bibitem{l2} {X. Tan, D.-W. Zhang, W. Zheng, X. Yang, S. Song, Z. Han,Y. Dong, Z. Wang, D. Lan, H. Yan, S.-L. Zhu, and Y. Yu, Experimental Observation of Tensor Monopoles with a Superconducting Qudit,}
\href{https://journals.aps.org/prl/abstract/10.1103/PhysRevLett.126.017702}{%
Phys. Rev. Lett. 126, 017702 (2021).}

\bibitem{l3} {D. L. Campbell, R. M. Price, A. Putra, A. Valdés-Curiel, D. Trypogeorgos, and I. B. Spielman, Magnetic phases of spin-1 spin-orbit-coupled Bose gases,} \href{https://www.nature.com/articles/ncomms10897}{%
Nat. Commun. 7, 10897 (2016).}

\bibitem{l4} {K. Sun, C. Qu, Y. Xu, Y. Zhang, and C. Zhang, Interacting spinorbit-coupled spin-1 Bose-Einstein condensates,} \href{https://journals.aps.org/pra/abstract/10.1103/PhysRevA.93.023615}{%
Phys. Rev. A 93, 023615 (2016).}

\bibitem{l5} {Z.-Q. Yu, Phase transitions and elementary excitations in spin-1 Bose gases with Raman-induced spin-orbit coupling,} \href{https://journals.aps.org/pra/abstract/10.1103/PhysRevA.93.033648}{%
Phys. Rev.A 93, 033648 (2016).}

\bibitem{l7} {E. J. König and J. H. Pixley, Quantum Field Theory of Nematic Transitions in Spin-Orbit-Coupled Spin-1 Polar Bosons,} \href{https://journals.aps.org/prl/abstract/10.1103/PhysRevLett.121.083402}{%
Phys. Rev. Lett. 121, 083402 (2018).}

\bibitem{l8} {I. Kuzmenko, T. Kuzmenko, Y. Avishai, and M. Sato, Spin-orbit coupling and topological states in an $F=\frac{2}{3}$ cold Fermi gas,} \href{https://journals.aps.org/prb/abstract/10.1103/PhysRevB.98.165139}{%
Phys. Rev. B 98, 165139 (2018).}


\bibitem{l9} {J. Hou, H. Hu, C. Zhang, Topological phases in pseudospin-$1$ Fermi gases with two-dimensional spin-orbit coupling,} \href{https://journals.aps.org/pra/abstract/10.1103/PhysRevA.101.053613}{%
Phys. Rev. A 101, 053613 (2020).}

\bibitem{l10} {G. Palumbo and N. Goldman, Revealing Tensor Monopoles through Quantum-Metric Measurements,} \href{https://journals.aps.org/prl/abstract/10.1103/PhysRevLett.121.170401}{%
Phys. Rev. Lett. 121, 170401 (2018).}

\bibitem{l11} {K. Murata, H. Saito, M. Ueda, Broken-axisymmetry phase of a spin-1 ferromagnetic Bose-Einstein condensate,} \href{https://journals.aps.org/pra/abstract/10.1103/PhysRevA.75.013607}{%
Phys. Rev. A 75, 013607 (2007).}

\bibitem{l12} {A. Lamacraft, Spin-1 microcondensate in a magnetic field,} \href{https://journals.aps.org/pra/abstract/10.1103/PhysRevA.83.033605}{%
Phys. Rev. A 83, 033605 (2011).}

\bibitem{sthe} {Y. Su, J. Hou, and C. Zhang, Universal intrinsic higher-rank spin tensor Hall effect,} \href{https://doi.org/10.1103/PhysRevB.107.085410}{%
Phys. Rev. B 107, 085410 (2023).}

\bibitem{f1} {M. Aidelsburger, M. Atala, and I. Bloch, Realization of the Hofstadter Hamiltonian with ultracold atoms in optical lattices,} \href{https://journals.aps.org/prl/abstract/10.1103/PhysRevLett.111.185301}{%
Phys. Rev. Lett. 111, 185301 (2013).}

\bibitem{f2} {S. Kolkowitz, S. L. Bromley, T. Bothwell, M. L. Wall, G. E. Marti, A. P. Koller, X. Zhang, A. M. Rey, and J. Ye, Spin–orbit-coupled fermions in an optical lattice clock,} \href{https://www.nature.com/articles/nature20811}{%
Nature 542, 66 (2017).}

\bibitem{scat1} C. Xu and J. E. Moore, Spin–orbit-coupled fermions in an optical lattice clock, \href{https://journals.aps.org/prb/abstract/10.1103/PhysRevB.73.045322}{%
Phys. Rev. B 73, 045322 (2006).}

\bibitem{scat2} J. Erhardt, M. Iannetti, F. Dominguez, E. M. Hankiewicz, B. Trauzettel, G. Profeta, D. D. Sante, G. Sangiovanni, S. Moser, R. Claessen, Backscattering in Topological Edge States Despite Time-Reversal Symmetry, \href{https://arxiv.org/abs/2503.11497}{%
arXiv: 2503.11497 (2025).}

\bibitem{w1} {A. A. Soluyanov and D. Vanderbilt, Computing Topological Invariants without Inversion Symmetry,} \href{https://journals.aps.org/prb/abstract/10.1103/PhysRevB.83.235401}{%
Phys. Rev. B 83, 235401 (2011).}

\bibitem{w2} {R. Yu, X. L. Qi, A. Bernevig, Z. Fang, and X. Dai, Equivalent Expression of Z2 Topological Invariant for Band Insulators Using the Non-Abelian Berry Connection,} \href{https://journals.aps.org/prb/abstract/10.1103/PhysRevB.84.075119}{%
Phys. Rev. B 84, 075119 (2011).}

\bibitem{sc1} {D. N. Sheng, Z. Y. Weng, L. Sheng, and F. D. M. Haldane, Quantum Spin-Hall Effect and Topologically Invariant Chern Numbers,} \href{https://journals.aps.org/prl/abstract/10.1103/PhysRevLett.97.036808}%
{Phys. Rev. Lett. 97, 036808 (2006).}

\bibitem{sc2} {E. Prodan, Robustness of the spin-Chern number,} \href{https://journals.aps.org/prb/abstract/10.1103/PhysRevB.80.125327}%
{Phys. Rev. B 80,125327 (2009).}

\bibitem{sc3} {H. Li, L. Sheng, D. Sheng and D. Y. Xing, Chern number of thin films of the topological insulator $Bi_2Se_3$,} \href{https://journals.aps.org/prb/abstract/10.1103/PhysRevB.82.165104}%
{Phys. Rev. B 82, 165104 (2010).}

\bibitem{cc5} {C. Zhang, Spin-orbit coupling and perpendicular Zeeman field for fermionic cold atoms: Observation of the intrinsic anomalous Hall effect,} \href{https://doi.org/10.1103/PhysRevA.82.021607}{
Phys. Rev. A 82, 021607(R) (2010).}

\bibitem{4} {G. Jotzu, M. Messer, R. Desbuquois, M. Lebrat, Th.Uehlinger, D. Greif, T. Esslinger, uperfluids, Experimental realization of the topological Haldane model with ultracold fermions, } \href{https://www.nature.com/articles/nature13915}%
{Nature 515, 237 (2014).}

\bibitem{QWZ} {M. C. Liang, Y. D. Wei, L. Zhang, X. J. Wang
, H. Zhang, W. W. Wang, W. Qi, X. J. Liu
, and X. Zhang, Realization of Qi-Wu-Zhang model in spin-orbit-coupled ultracold fermions,} \href{https://doi.org/10.1103/PhysRevResearch.5.L012006}%
{Phys. Rev. Research 5, L012006 (2023).}

\bibitem{cc1} {M. C. Beeler, R. A. Williams, K. Jiménez-García, L. J. LeBlanc, A. R. Perry and I. B. Spielman, The spin Hall effect in a quantum gas,} \href{https://doi.org/10.1038/nature12185}{
Nature 498, 201 (2013).}

\bibitem{cc2} {S. L. Zhu, H. Fu, C. J. Wu, S. C. Zhang, and L. M. Duan, Spin Hall Effects for Cold Atoms in a Light-Induced Gauge Potential,} \href{DOI: https://doi.org/10.1103/PhysRevLett.97.240401}{
Phys. Rev. Lett. 97, 240401(2006).}

\bibitem{cc3} {C. J. Kennedy, G. A. Siviloglou, H. Miyake, W. Cody Burton, and W. Ketterle, Spin-Orbit Coupling and Quantum Spin Hall Effect for Neutral Atoms without Spin Flips,} \href{https://doi.org/10.1103/PhysRevLett.111.225301}{
Phys. Rev. Lett. 111, 225301 (2013).}

\bibitem{cc4} {Q. X. Lv, Y. X. Du, Z. T. Liang, H. Z. Liu, J. H. Liang, L. Q. Chen, L. M. Zhou, S. C. Zhang, D. W. Zhang, B. Q. Ai, H. Yan, and S. L. Zhu, Measurement of Spin Chern Numbers in Quantum Simulated Topological Insulators,} \href{https://doi.org/10.1103/PhysRevLett.127.136802}{
Phys. Rev. Lett. 127, 136802 (2021).}



\bibitem{ll1} {H. Y. Xu, Y. C. Lai, Superscattering of a pseudospin-1 wave in a photonic lattice, } \href{https://journals.aps.org/pra/abstract/10.1103/PhysRevA.95.012119}{
Phys. Rev. A 95, 012119 (2017).}


\bibitem{ll2} {X. Luo, L. Wu, J. Chen, Q. Guan, K. Gao, Z. F. Xu, L. You, and R. Wang, Tunable atomic spin-orbit coupling synthesized with a modulating gradient magnetic field,}
\href{https://doi.org/10.1038/srep18983}{Sci. Rep. 6, 18983 (2016).}


\bibitem{ll3} {S. Taie, Y. Takasu, S. Sugawa, R. Yamazaki, T. Tsujimoto, R. Murakami, and Y. Takahashi, Realization of a SU(2)×SU(6) System of Fermions in a Cold Atomic Gas,} \href{https://journals.aps.org/prl/abstract/10.1103/PhysRevLett.105.190401}%
{Phys. Rev. Lett. 105, 190401 (2010).}

\bibitem{ll4} {X. Zhang, M. Bishof, S. L. Bromley, C. V. Kraus, M. S. Safronova, P. Zoller, A. M. Rey, and J. Ye, Spectroscopic Observation of SU(N)-Symmetric Interactions in Sr Orbital Magnetism,} \href{https://www.science.org/doi/10.1126/science.1254978}%
{Science 345, 1467 (2014).}

\bibitem{ll5} {H. Y. Xu, Y. C. Lai, Revival resonant scattering, perfect caustics, and isotropic transport of pseudospin-1 particles,} \href{https://journals.aps.org/prb/abstract/10.1103/PhysRevB.94.165405}%
{Phys. Rev. B 94, 165405 (2016).}

\bibitem{h1} {L. Tarruell, D. Greif, T. Uehlinger, G. Jotzu, and T. Esslinger, Creating, Moving and Merging Dirac Points with a Fermi Gas in a Tunable Honeycomb Lattice, } \href{https://www.nature.com/articles/nature10871}%
{Nature 483, 302 (2012).}

\bibitem{Bloch16} {N. Fläschner, B. S. Rem, M. Tarnowski, D. Vogel, D.-S. Lühmann, K. Sengstock, and C. Weitenberg, Experimental reconstruction of the Berry curvature in a Floquet Bloch band,} \href{https://doi.org/10.1126/science.aad4568}%
{Science 352, 1091 (2016).}

\bibitem{Schneider16} {T. Li, L. Duca, M. Reitter, F. Grusdt, E. Demler, M. Endres, M. Schleier-Smith, I. Bloch, and U. Schneider, Bloch state tomography using Wilson lines,} \href{https://doi.org/10.1126/science.aad5812}%
{Science 352, 1094 (2016).}

\bibitem{h3} {K. Wintersperger, C. Braun, F. N. Ünal, A. Eckardt, M. D. Liberto, N. Goldman, I. Bloch, and M. Aidelsburger, Realization of an anomalous Floquet topological system with ultracold atoms,} \href{https://doi.org/10.1038/s41567-020-0949-y}%
{Nat. Phys. 16, 1058–1063 (2020).}


\bibitem{stm1} {H. Hu, J. Hou, F. Zhang, and C. Zhang, Topological Triply Degenerate Points Induced by Spin-Tensor-Momentum Couplings,} \href{https://journals.aps.org/prl/abstract/10.1103/PhysRevLett.120.240401}%
{Phys. Rev. Lett. 120, 240401 (2018).}

\bibitem{stm2} {X. W. Luo, K. Sun, and C. Zhang, Spin-Tensor–Momentum Coupled Bose-Einstein Condensates,} \href{https://journals.aps.org/prl/abstract/10.1103/PhysRevLett.119.193001}%
{Phys. Rev. Lett. 119,193001 (2017).}


\end{thebibliography}
\end{document}